\documentclass[fleqn,10pt]{wlscirep}
\title{On Anderson Localization and Chiral Anomaly in Disordered Time-Reversal Invariant Weyl Semimetals: Nonperturbative and Berry Phase Effects}

\author[1,2,*]{Imam Makhfudz}

\affil[1]{Laboratoire de Physique Th\'{e}orique--IRSAMC, CNRS and Universit\'{e} de Toulouse, UPS, F-31062 Toulouse, France}
\affil[2]{Universit\'{e} Lyon, ENS de Lyon, Universit\'{e} Claude Bernard, CNRS, Laboratoire de Physique, F-69342 Lyon, France}

\affil[*]{imam.makhfudz@ens-lyon.fr}

\begin{abstract}
Weyl semimetal, a three-dimensional electronic system with relativistic linear energy dispersion around gapless points carrying nontrivial Berry charge, is predicted to exhibit a wealth of unique response and transport properties.A crucial question is whether those properties are robust against disorder and whether Anderson localization occurs.In this work, the effects of nonperturbative topological (vortex loop) excitations and Berry phase in disordered time-reversal invariant 3d Weyl semimetal are studied.It is shown that the chiral symmetry is restored in the nonlinear sigma model describing the diffusons upon disorder average as any net topological term and its delocalization result do not take effect at sufficiently short length scales.Anderson localization occurs at sufficiently strong disorder and we predict that chirality and related phenomena disappear at such transition.Nevertheless, we uncover a mechanism that originates from Berry phase that impedes such localization effect.We show the occurrence of destructive interference between the vortex loops and between scattering paths due to the the vortex loops' Berry phase which resists the Anderson localization.We emphasize the applicability of our theory to the candidate Weyl materials where we point out the consistency of our theory with a recent experimental finding of the absent chiral anomaly in a noncentrosymmetric Weyl semimetal. 
\end{abstract}
\begin{document}

\flushbottom
\maketitle
%
%
\thispagestyle{empty}

\section*{Introduction}

Weyl semimetal (WSM) is a new class of three dimensional materials with topologically protected gapless isolated points as its Fermi surface with relativistic linear dispersion spectrum and well-defined chirality for each Weyl point.The concept of Weyl fermions arose in the context of high energy (elementary particle) physics from the solutions of the massless Dirac equation with definite chirality.According to Berry phase theory \cite{Berry}, the Weyl points are Berry charge monopoles in momentum space \cite{Volovik} \cite {NiuRMP} with opposite signs for Weyl points of opposite chiralities.The topological protection arises from the fact that the monopole charges of opposite signs can only be created or annihilated in pairs.WSM is characterized by unique response properties in the form of chiral magnetic effect arising from the phenomenon of axial anomaly between two Weyl points of opposite chiralities, surface Fermi arc \cite{Ashvin}, and anomalous Hall effect \cite{AHEinWeyl} which gives rise to new transport properties \cite{HosurQi}.In realistic situations, candidate materials hosting Weyl points always contain a finite amount of disorder.Recent theoretical and numerical studies of disordered WSM uncovered a number of new phases with interesting transport and critical properties.One of the important issues is the occurrence and nature of Anderson localization \cite{Anderson1958} in this system.

The chirality of Weyl fermion is the special feature of this relativistic fermion, which distinguishes it from the chirally symmetric Dirac fermion.This chirality has been argued to give rise to topological protection mechanism against Anderson localization in disordered WSM in unitary class, where time-reversal symmetry is broken, via the presence of topological $\theta$ term accompanying the nonlinear sigma model (NLSM) \cite{AltlandBagrets}, in analogy to the physics of quantum Hall effect \cite{Pruisken}\cite{ZiqiangWang}.This protection mechanism is effective in the large distance scales valid in the presence of sufficiently smooth disorder or disorder with long enough range, such that the internode scattering can be neglected.Renormalization group (RG) study \cite{GoswamiSudip}\cite{Gurarie} and numerical calculations \cite{Sarma} indeed observed robust conducting state at weak disorder, in the form of `diffusive metallic' state \cite{Fradkin1986}.This was also confirmed by Boltzman equation and self-consistent Born approximation calculations \cite{BiswasRyu}\cite{Japanesepaper}\cite{Brouwer}.However, numerics also found insulating Anderson localized phase at strong enough disorder \cite{Sarma} in model with pair(s) of Weyl points in time-reversal invariant WSM.It is well known that the density of states vanishes quadratically at the Weyl point energy, making the semimetal robust against weak disorder.But as disorder increases, it eventually induces finite density of states above above a critical disorder strength \cite{Sarma} and allows for more particle scatterings.Internode backscattering between Weyl points of opposite chiralities further strengthens the tendency towards localization.

In this paper, we will first demonstrate using supersymmetric method that the chirality of WSM does not appear in the NLSM upon averaging over disorder, giving rise to the field theory of effectively chirally symmetric disordered Dirac fermion system.We show this both in the case of  independent Weyl nodes as well as that with node mixing.This `chiral symmetry recovery' means the delocalization effects of any net topological $\theta$ term of paired chiral Weyl points disappear at short enough length scales when the two Weyl points are connected by internode scattering and the $\theta$ terms from the two nodes cancel out.Then, using replica formulation of NLSM, we analyze the nonperturbative effect in the form of vortex loop proliferation on the disordered WSM which provides another mechanism that drives localization.We next consider the effects of the Weyl points' intrinsic Berry phase as well as that of the vortex loops on the localization physics.We will show that while disorder-induced density of states, intercone backscattering, and vortex loop proliferation promote Anderson localization, the Berry phase effect of the vortex loops surprisingly impedes the localization due to the induced destructive interference between the vortex loops and between the particle scattering paths.This mechanism is the key to the robustness of metallicity in WSM, in the diffusive metallic state \cite{Sarma}.We consider time-reversal invariant (TRI) WSM with chirality coming from broken inversion symmetry (parity).Our theory applies both to Weyl systems without spin-orbit coupling, such as that which has been predicted to occur in real material \cite{HuangPNAS} as well as pioneering candidate material such as TaAs where the Weyl points do require spin-orbit coupling \cite{HuangNatcomm}\cite{WengPRX}\cite{XuScience}\cite{LvPRX}\cite{YangNatphys}.The absence of inversion symmetry naturally gives rise to Berry phase effect \cite{NiuRMP}.Finally, we show that the chiral symmetry recovery is a consistent explanation for the recent observed absence of chiral anomaly in a noncentrosymmetric WSM. 

\section*{Results}

\subsection*{Symmetry and Field Theory:Chiral Symmetry Recovery in Disordered WSM}

We will show that chiral symmetry, broken in the clean TRI WSM, can be recovered upon averaging over (TRI scalar potential) disorder.To this end, we consider single-particle low-energy Hamiltonian for WSM in the presence of general 4-vector (scalar, 3-vector) potential disorder, is given by

\begin{equation}\label{WSMhamiltonian}
H=\tau^z v_F \mathbf{\sigma}\cdot\mathbf{k}+v_F\sigma\cdot\mathbf{k}_0 +\mu+ \sigma_{\mu}V_{\mu}(\mathbf{r})
\end{equation}
written in basis $(\psi_R,\psi_L)^T$ with $\psi_{\alpha=R,L}=(\psi_{\alpha,+},\psi_{\alpha,-})^T$,  which describes a pair of Weyl points at energy-momentum $\pm\mathbf{k}_0$ with energy position $\epsilon=\pm\mu$ relative to Fermi energy $\epsilon=0$ at half-filling.The $\sigma$ is Pauli matrix representing some sublattice degree of freedom, $\tau^z$ is the $z$ Pauli matrix acting in the node (right R-left L) space, corresponding to the chirality space, with eigenvalues $\eta_{R/L}=\pm 1$ representing the chirality of the Weyl point, $V_{\mu}$ represents the scalar potential ($V_0$), vector potential ($V_x$ and $V_y$), and mass disorders, which are white noise and Gaussian correlated.The presence of $\tau^z$ explicitly breaks the chiral symmetry generated by the chirality operator $Ch=\tau^x$.Since chirality operator can be written as a product of the time-reversal and particle-hole  operators; $Ch=\mathcal{T}\times \Xi$, a chiral system such as WSM can therefore be obtained by breaking either $\mathcal{T}$ or $\Xi$.We consider TRI WSMs without inversion symmetry (parity), which gives $\Xi$-breaking band structure.This is realized in appropriate noncentrosymmetric Weyl materials such as SrSi$_2$ \cite{HuangPNAS} and modeled by the Hamiltonian Eq.(\ref{WSMhamiltonian}) which has an effective time-reversal symmetry $T_{\perp}$; $H=\sigma_yH^*\sigma_y$ \cite{OGMirlin}.It is this $T_{\perp}$ that we take to define the time-reversal invariance of $H$.Since $V_x,V_y$, and $V_z$ all break $T_{\perp}$, we only consider scalar potential disorder $V_0$ with $\langle V^2_0\rangle=1/(2\pi \nu_{\mathrm{DOS}}\tau)$ where $\nu_{\mathrm{DOS}}$ is the density of states at the Fermi energy and $\tau$ is the mean free (relaxation) time.The shift $\mu$ is very small relative to the energy scale $E$ that defines the bandwidth of energy below which the linear energy dispersion description is valid.Since $\mu$ is very small, the Weyl points are still very close to the Fermi energy, so that practically the clean system is semimetallic at half-filling.   

We will first consider long-range disorder where internode mixing can be neglected, allowing us to focus on one node with a given chirality $\eta_{R/L}=\pm 1$.We use supersymmetric method \cite{Efetov} by defining an 8-spinor with elements coming from the sublattice, retarded-advanced (R-A), and boson-fermion (B-F) spaces.Taking average over disorder, we obtain effective fermion action with 4-fermion term generated by the disorder.Upon performing the Hubbard-Stratonovich transformation in terms of the supermatrix Q, we obtain
\begin{equation}
S[\overline{\psi}_R,\psi_R]=\int d^3r \left[-i\overline{\psi}_R \left(-\eta_Rv_F\sigma\cdot i\nabla+\mu+v_F\sigma\cdot\mathbf{k}_0 +\frac{(\omega+i\delta)}{2}\Lambda-\varepsilon+\frac{\omega}{2}+\frac{i}{2\tau}Q\right) \psi_R
+ \frac{\pi\nu_{\mathrm{DOS}}}{8\tau} \mathrm{Str}Q^2\right] 
\end{equation}
where $\varepsilon$ is energy (frequency), and $\Lambda=\mathrm{diag}(1,-1)$ is diagonal matrix in the R-A space, the supertrace $\mathrm{Str}$ involves the sums over sublattice, R-A, and B-F spaces, and we take right-handed Weyl fermion with $\eta_R=1$ for definiteness.The $\left((\omega+i\delta)/2\right) \Lambda$ breaks the symmetry in the R-A space.Now, we integrate out the fermions and obtain
\begin{equation}\label{effaction}
S[Q]=\int \left[-\frac{1}{2}\mathrm{Str}\mathrm{log}\left(-iH_R-i\frac{(\omega+i\delta)}{2}\Lambda+\frac{Q(\mathbf{r})}{2\tau}\right)+\frac{\pi\nu_{\mathrm{DOS}}}{8\tau}\mathrm{Str}Q^2(\mathbf{r})\right]d^3 r
\end{equation}
where $H_R=\eta_Rv_F\sigma\cdot(-i\nabla)+\mu+v_F\sigma\cdot\mathbf{k}_0-\varepsilon+\omega/2$.

We then take mean field approximation around the saddle point describing the broken R-A state.Performing gradient expansion, we obtain NLSM
\begin{equation}\label{NLSM}
S^{\mathrm{NLSM}}_{\mathrm{SUSY}}[Q]=K_{\mathrm{SUSY}}\int d^3r\mathrm{Str}\left[(\nabla Q)^2\right]
\end{equation}
with
\begin{equation}
K_{\mathrm{SUSY}}=(\Sigma^2+3\Delta^2)/(48\pi v_F\Delta)
\end{equation}
where $\Delta=1/(2\tau)$, $\Sigma=\sqrt{(\varepsilon-\mu)^2+v^2_F|\mathbf{k}_0|^2}$ and $Q$ describes the diffusons; the Goldstone modes of the R-A symmetry breaking and subject to the constraint $Q^2=\mathbb{I}$.We note that the chirality factor $\eta_R=1$ is no longer present in the effective NLSM, as it comes up as $\eta^2_R$.Derivation for $\eta_L=-1$ gives the same NLSM.This means the chiral symmetry broken in the clean WSM is recovered in dirty WSM, upon disorder averaging.We thus have emergent chiral symmetry in the presence of scalar potential disorder.Remarkably, we found that this `chiral symmetry recovery' holds even in the presence of internode scattering applicable to short-range correlated disorders, in a full treatment of paired Weyl points problem \cite{SuppMat}.

We explain the meaning of the above result by employing renormalization group (RG) approach which is a method to derive a low-energy effective theory by successively integrating out higher energy degrees of freedom \cite{WilsonRMP}.For our time-reversal invariant system within nonlinear sigma model formalism \cite{Hikami}, we obtain RG equation for the longitudinal conductance $g_{xx}$ in general form

\begin{equation}\label{RGlongitudconductivityperturbative}
\frac{dg_{xx}}{dl}=g_{xx}-{g^0_{xx}}^*
\end{equation}  
where ${g^0_{xx}}^*$ is a nonzero positive constant, corresponding to a critical conductance.The first term on the right hand side comes from the length scale rescaling $\epsilon g_{xx}$ with $\epsilon=d-2=1$ for $d=3$, and the second term is a localization effect.This RG equation has nontrivial fixed point at ${g^0_{xx}}^*$ corresponding to a critical disorder strength.For disorder strengths weaker than the critical value, we have $g^0_{xx}>{g^0_{xx}}^*$ and the system flows to the semimetallic (or diffusive metallic) $g_{xx}\rightarrow\infty$ fixed point.As one increases the disorder, $g^0_{xx}$ decreases and above a critical disorder, $g^0_{xx}<{g^0_{xx}}^*$ so that the system flows to the Anderson insulating state $g_{xx}\rightarrow 0$.To analyze the topological property, we consider the topological term accompanying the above NLSM; 

\begin{equation}\label{topologicalthetaterm}
S_{\theta}[Q]=\frac{\theta}{8\pi}\int d^3x \mathrm{Str}\left[Q\nabla_{x}Q\nabla_{y}Q\right]
\end{equation} 
where $\theta=\sum_{a=R,L}(2\pi\sigma^a_{xy}+\theta^a_{\mathrm{Berry}})$ with Hall conductivity $\sigma^a_{xy}$ and intrinsic Berry phase $\theta^a_{\mathrm{Berry}}$ of each Weyl node ($\sigma^a_{xy}=0$, $\sum_{a=R,L}\theta^a_{\mathrm{Berry}}$ is vanishingly small in our system, due to the parity breaking from small asymmetry in the energy dispersion caused by the the energy shift $\mu$).A crucial point is that, unlike in 2d, in 3d the integral in $S_{\theta}[Q]$ is not an exact topological invariant (defined by homotopy group $\pi_d(G/H)$ of the coset space $G/H$ described later), but is nevertheless referred to as topological term due to its form.Furthermore, any $\theta/(2\pi)$ taking value away from quantized (integer) value will be renormalized by nonperturbative instanton effect \cite{LevineLibbyPruisken}\cite{Pruiskendiluteinstanton} to the nearest integer value; to zero in our theory.This is the case even in 2d quantum Hall effect problem, where the $\theta$ term (the Pruisken term) \cite{Pruisken} is an exact topological invariant corresponding to homotopy group $\pi_2(G/H)$.We will show that there is direct analog of instanton effect in our 3d system.It is the vortex loop excitations, which should be responsible for the the down renormalization of the topological $\theta$ term.This $\theta\rightarrow 0$ flow is accompanied by conductivity flow $\sigma_{xx}\rightarrow 0$ (corresponding to the finite conductance ${g^0_{xx}}^*=\sigma_{xx}L$ as we take the length scale (system size )$L=\exp(l)\rightarrow \infty$) flow \cite{Pruiskendiluteinstanton}, which is fully consistent with numerical calculation \cite{Sarma} that found that typical density of states $\rho_t=0$ and thus $\sigma_{xx}=0$ in the Anderson insulating state.In addition, according to the hierarchical classification and analysis of topological terms \cite{RyuNJP}, one can have a $\mathbb{Z}_2$ version of Eq.(\ref{topologicalthetaterm}) for TRI system, but it lives on the 2d surface of our 3d system.In addition, there is also Chern-Simons term $S_{CS}[A]\equiv \mathrm{Tr}[AdA+\frac{2}{3}A^3]$ where $A_i=T^{-1}\partial_iT, i=x,y,z$ and $T\in G/H$ \cite{RyuNJP} is diffuson matrix field related to the matrix $Q$ via $Q=T\tau^{R-A}_3 T^{-1}$, $\tau^{R-A}_3$ the third Pauli matrix in retarded-advanced space, living in appropriate coset space $G/H$ (to be described in the next section).This Chern-Simons term however carries opposite signs between the two Weyl points of opposite chiralities $S^{n}_{CS}[A_n]=(-1)^nS_{CS}[A_n]$ with node index $n=1,2$ \cite{AltlandBagrets} so that their effect cancel out when the two Weyl points are connected by scattering\cite{Cancellation}.Furthermore, this Chern-Simons term does not contribute to any renormalization since its overall coefficient is numerical constant.None of these terms can therefore give rise to protection against Anderson localization in the \textit{bulk} at short enough distance scales, where internode scattering is inevitable.Our system of a pair of Weyl points under strong disorder thus behaves like a pair of Dirac points, allowed by symmetry\cite{NagaosaNatComm}.

To relate this result to the topological response property, especially with regard to the fascinating chiral anomaly in Weyl fermion system \cite{NielsenNinomiya}, we couple the diffuson to the electromagnetic field by Peierls substitution, $\nabla_{\mu}\rightarrow\nabla_{\mu}+ieA_{\mu}$ in $S_{\theta}[Q]$.Integrating out the diffusons, one obtains an axion term $S_{\mathrm{axion}}[A]=(e^2/8\pi^2)\int d^{3+1}x\theta\epsilon^{\mu\nu\lambda\delta}\partial_{\mu}A_{\nu}\partial_{\lambda}A_{\delta}$ that describes chiral anomaly.It is thus clear that since $\theta$ renormalizes to zero, the chiral anomaly also perishes at the Anderson localization transition.This is the direct physical consequence of the chiral symmetry recovery.The above analysis shows that 1. Anderson localization occurs in Weyl semimetal at sufficiently strong disorder 2. The chirality and the associate phenomena disappear at the Anderson transition.Verification of the second point thus implies the first one.

\subsection*{Nonperturbative Effects in Disordered 3d WSM:Berry Phase and Vortex Loop Excitations}

We have considered thus far the perturbative influence of the disorder plus its topological terms, which do not help to counter Anderson localization in the bulk in our TRI Weyl system.Inspired by the profound significance of nonperturbative effect in quantum field theory \cite{InstantonQFT} as well as that in condensed matter, as illustrated in the study of Anderson localization in 2d Dirac fermion system and topological insulator \cite{Konig}\cite{Ryu}\cite{FuKane}, we consider such effect in the field theory description of Anderson localization in disordered 3d WSM.In this case, the topological excitations describing this effect take the form of vortex line defects which in the lowest energy state will be closed vortex line, called vortex loop or vortex ring.Generally speaking, topological excitations tend to disorder the system, restore the broken symmetry, and drive the system into the insulating localized state.  

In this work, we point out a mechanism accompanying such vortex loops which however counters the tendency towards localization and leads to the robustness of WSM against going into Anderson insulating phase; the Berry phase effect, by making analogy with spin systems \cite{TanakaTotsukaHu}\cite{IM-PPprb}.Heuristic argument: Minimal model of WSM contains a pair of Weyl points with opposite chiralities, corresponding to Berry monopole charges of opposite signs.The net Berry phase on surface enclosing this pair of Weyl points is therefore zero which should give rise to constructive quantum interference between the scattering trajectories in Anderson localization \cite{Anderson1958} and also between the vortex loops that can arise in the NLSM description of the disordered WSM.However, the vortex loops themselves may posses nonzero Berry phase, which in general takes values away from integer multiples of $2\pi$ and gives rise to destructive interference between the vortex loops and also between the particle scattering paths.This destructive interference impedes the tendency for localization.   

To describe the above effect, we consider the NLSM field theory again, this time within replica formalism.The NLSM takes the form $S^{\mathrm{NLSM}}_{\mathrm{replica}}=K\int d^3r \mathrm{Tr}[(\nabla Q(\mathbf{r}))^2]$, where $K$ represents longitudinal conductivity.Our formulation will be in terms of system with time-reversal and spin rotational symmetries and thus belongs to the Wigner-Dyson orthogonal class AI \cite{Schnyder}, which is not much considered in literature.Importantly however, our theory and its results will still be applicable to the more familiar symplectic class AII, where time-reversal symmetry is preserved but spin-rotational symmetry is broken, due to spin-orbit coupling for example.The reason is in such system, the 4-band Hamiltonian exemplified by Eq.(\ref{WSMhamiltonian}) can be reduced to 2-band Hamiltonian for each spin (or pseudospin, played by sublattice in our model) sector, each of which hosts its own Weyl points \cite{Ojanen}.Our result therefore applies to Weyl points for both of spin sectors.For the orthogonal class AI, the manifold (coset space) of the fermionic replica NLSM is $G/H=Sp(2N)/(Sp(N)\times Sp(N))$, where $Sp(\cdots)$ is symplectic matrix group, $N$ is the number of retarded or advanced replicas \cite{SchnyderAIP}.We found that the $2N\times 2N$ matrix $Q(\mathbf{r})$ can be written in such a way that exposes a $U(1)$ gauge structure as follows

\begin{equation}\label{Qabelian}
Q(\mathbf{r})=\left( \begin{array}{cc}
e^{i\phi(\mathbf{r})}\tilde{Q}_{11}(\mathbf{r})&e^{-i\phi(\mathbf{r})}\tilde{Q}_{12}(\mathbf{r})\\
e^{i\phi(\mathbf{r})}\tilde{Q}_{21}(\mathbf{r})&e^{-i\phi(\mathbf{r})}\tilde{Q}_{22}(\mathbf{r})
\end{array} \right)
\end{equation}
where the $\tilde{Q}_{ab}(\mathbf{r})$ is each $N\times N$ matrix, subject to appropriate constraints \cite{SuppMat}.Such `Abelian content extraction' of non-Abelian theory is common in quantum field theory and amounts to a form of gauge fixing to manifest relevant physical degree of freedom \cite{Hooft} represented by the phase angle field $\phi(\mathbf{r})$.NLSM does have such a gauge degree of freedom \cite{BrezinHikamiZinn}. 

In analogy with studies in fractional quantum Hall effect and superfluids \cite{BerryFQHEsuperfluids}, we obtain the Berry phase action in terms of the diffuson matrix field $Q(\mathbf{r})$ as

\begin{equation}\label{Berryaction}
S_B=\int d^3r\oint_C d\mathbf{r}'\cdot \mathrm{Tr}[-iQ^{-1}(\mathbf{r}'-\mathbf{r})\nabla_{\mathbf{r}'-\mathbf{r}} Q(\mathbf{r}'-\mathbf{r})]\rho(\mathbf{r})
\end{equation}
where $C$ is a closed contour in real space, $\rho(\mathbf{r})$ is the density of the diffusons.For WSMs with paired Weyl points, as noted earlier, the net intrinsic Berry phase due to broken inversion symmetry is nonzero but vanishingly small.The Berry action $S_B$ will be dominated by vortex loop contribution computed as follows.

\begin{figure}
 \includegraphics[scale=0.20]{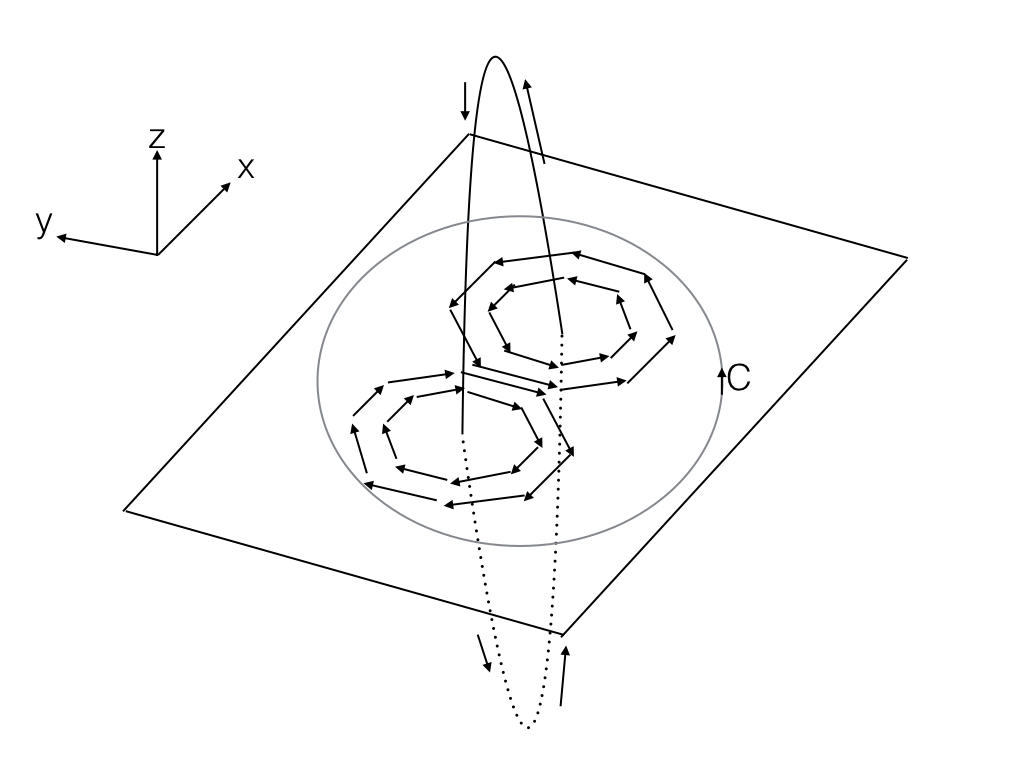}
 \label{fig:VortexLoop}
 \caption{A single vortex loop in 3d space, piercing the 2d $x-y$ plane giving rise to a pair of vortices with opposite circulations and Berry phase evaluated along contour $C$.}
\end{figure} 

Since symplectic matrices form a Lie group, one can represent the matrix $Q$ in terms of the generators $T^a$ of the associated Lie algebra; $Q=\mathbb{I}+\sum^{D-1}_{a=1}Q_aT^a$ where $D$ is the dimension of the coset space $G/H$.Then, a configuration of matrix field $Q(\mathbf{r})$ that hosts topological defect can be chosen, which in the simplest nontrivial case is given by \cite{SuppMat},

\begin{equation}\label{Qvortexloop}
Q(\mathbf{r})=\mathbb{I}+|p\rangle \left( \begin{array}{cc}
(1+i)(e^{i\phi(\mathbf{r})}-1)&e^{-i\phi(\mathbf{r})}-1\\
e^{i\phi(\mathbf{r})}-1&(1-i)(e^{-i\phi(\mathbf{r})}-1)
\end{array} \right)\langle p|
\end{equation} 
where $\mathbb{I}$ is $2N\times 2N$ identity matrix, with $N$ the number of retarded or advanced replicas and $\phi(\mathbf{r})$ describes the `projection' of the vortex loop field on $x-y$ plane at fixed $z=0$.In Eq.(\ref{Qvortexloop}), $|p\rangle$ represents a unit ($2N$-column) vector in replica space, where its $p^{\mathrm{th}}$ element is unity while all the other elements are zero.Using Eqs. (\ref{Berryaction}) and (\ref{Qvortexloop}) and considering a vortex line with flux line vector $\mathbf{r}_v$ giving $\rho(\mathbf{r})=\delta(\mathbf{r}-\mathbf{r}_v)$, we obtain for the Berry action

\[
S_B=-i\int d\varphi \frac{\partial \phi(R,\varphi)}{\partial \varphi}\frac{-2\cos\phi(R,\varphi)}{1-2\sin \phi(R,\varphi)}
\]
\begin{equation}\label{vortexloopBerryphase}
=-i\mathrm{log}\left|\frac{1-2\sin\phi_f}{1-2\sin\phi_i}\right|=-i\phi_1
\end{equation}
where we have taken, as the contour, a circle of radius $R=|\mathbf{r}-\mathbf{r}_v|$ on $x-y$ plane described by azimuthal angle $\varphi$.A single vortex loop is modeled in terms of the `projection' of its flux line on the $x-y$ plane, assuming a (static) vortex loop `stretched' along $z$ direction so that it `pierces' the $x-y$ plane in direction normal to the plane (parallel to the $z$ axis), giving rise to a pair of vortices with opposite circulations but with the same winding number.This is illustrated in Fig. 1.In this case, we can set $\phi_{i(f)}=\phi(R\rightarrow \infty,\varphi=-\pi/2\pm 0^+)$ with $\varphi$ measured from the $x$ axis counter clockwise.

Analyzing Eq.(\ref{vortexloopBerryphase}), the Berry phase of the vortex loop is always zero unless the vortex field pattern $\phi(\mathbf{r})$ breaks inversion symmetry, or breaking the mirror symmetry with respect to reflections about $x$ and $y$ axes.Since our WSM breaks the real space inversion symmetry, it is clearly intuitive that the latter condition is naturally satisfied because highly symmetric vortex field will cost higher energy in an inversion asymmetric system and will not be preferred.This is a remarkable observation that seals the consistency of our theory.This result suffices to establish that a) The Berry phase $S_B$ of a vortex loop is nonzero and b) in general is not integer multiple of $2\pi$, and thus tends to give rise to destructive interference between the vortex loops themselves as well as between the particle scattering trajectories, since the resulting partition function $Z_{vl}\sim \exp(-S_B)\neq 1$.In passing, we note that the Berry phase of \textit{open} vortex line, where $\phi_i=\phi_f$, is always $0$ ($\equiv 2\pi$), consistent with the time-reversal symmetry of our system.

\subsection*{Vortex Loop Proliferation and Correction of Longitudinal Conductance}

This work emphasizes that nonperturbative effect in the form of topological vortex loop excitations has to be taken into account in the description of disordered Weyl semimetal as one goes beyond the regime of validity of perturbation theory.For the vortex loops to be of any relevance however, we have to verify that they are able to proliferate and this can be analyzed by considering vortex loop fugacity, that we now investigate.To this end, we substitute the expression for the matrix field $Q(\mathbf{r})$ for AI class in Eq.(\ref{Qabelian}) into the expression for the NLSM $S^{\mathrm{NLSM}}_{\mathrm{replica}}=K\int d^3 r \mathrm{Tr}[(\nabla Q(\mathbf{r}))^2]$.We obtain
\begin{equation}\label{3dXYmodel}
S^{\mathrm{XY}}=K'\int d^3 r(\nabla \phi(\mathbf{r}))^2
\end{equation}
where $K'=\eta K$ with $\eta$ is a proportionality constant determined by the mean-field configuration of the block matrix elements of the full symplectic matrix field $Q(\mathbf{r})$ describing the diffusons \cite{SuppMat}.We note that the resulting NLSM takes the form of the familiar 3d XY model.Therefore, we can treat the NLSM of our system using 3d XY model. 

With this, we can use the result of RG study of 3d XY model, which gives the following RG equations for $J=2 K'$ representing the inter-vortex loop interaction coupling and the vortex loop fugacity $\nu$ \cite{3dXYmodelRG}.

\begin{equation}
\frac{dJ}{dl}=J-\frac{2\pi^3}{3}\nu J^2,\frac{d\nu}{dl}=\nu\left[6-\frac{\pi^2}{2}J\left(1+\mathrm{log}\frac{a}{a_c}\right)\right]
\end{equation}
These RG equations have nontrivial fixed point with finite vortex loop fugacity, located at $\nu^*=(1+\mathrm{log}(a/a_c))/(8\pi)$ and $J^*=12/(\pi^2(1+\mathrm{log}(a/a_c)))$ where $a$ is the short-distance cutoff (e.g. lattice spacing) and $a_c$ is the size of the vortex loop core, compared to 2d XY model where the nontrivial fixed point occurs at zero fugacity, which means vortices can start proliferate at arbitrarily small fugacity, describing the well-known BKT transition \cite{BKT}.On the other hand in 3d WSM, one needs a finite concentration of vortex loops in order for them to proliferate and drive localization; the vortex loop-driven localization effect onsets only after enough number of vortex loops accumulate.The WSM is known to display diffusive metal phase \cite{Fradkin1986} described precisely by the NLSM discussed here.This phase exists over extended regime of the phase diagram as function of the disorder \cite{Sarma}.Beyond perturbative mechanisms, the Berry phase effect of vortex loops strengthen the robustness of such diffusive metal phase, before Anderson localization prevails. 
\begin{figure}
\includegraphics[scale=0.20]{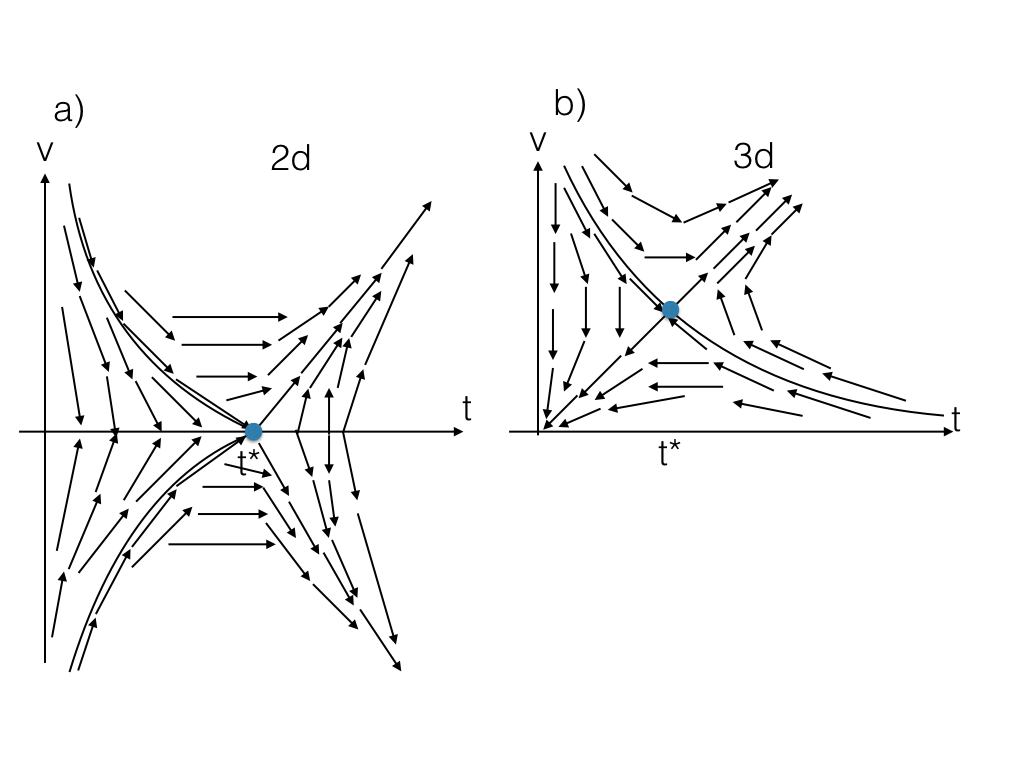}
 \includegraphics[scale=0.20]{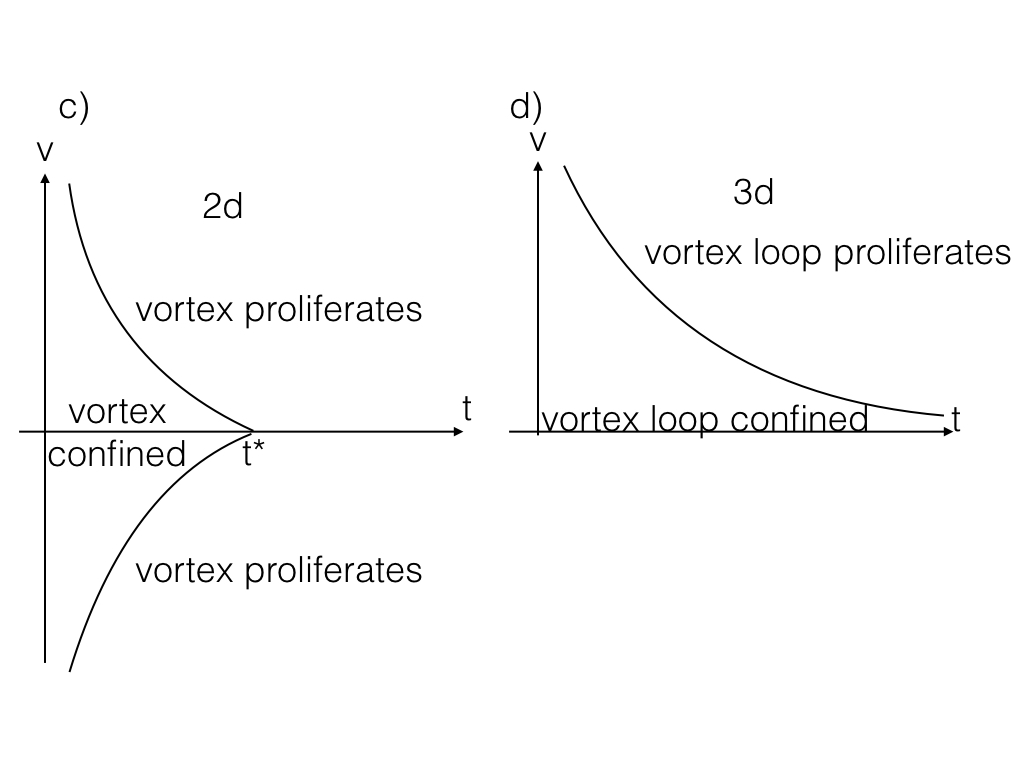}
 \label{fig:VortexLoop}
 \caption{The RG flow and disorder strength-fugacity $t-\nu$ phase diagrams of the proliferation of a)c) vortex of 2d XY model in 2d symplectic $\mathbb{Z}_2$ TI and b)d) vortex loop of 3d XY model in 3d orthogonal TRI WSM. }
\end{figure} 

As comparison, in 2d symplectic $\mathbb{Z}_2$ topological insulator (TI), $\epsilon=1-N$ RG analysis of $G/H=O(2N)/(O(N)\times O(N))$ NLSM ($\equiv$ 2d XY model at $\epsilon=0$) \cite{FuKane}, where $O(\cdots)$ represents orthogonal matrix group, found nontrivial fixed points $(\Delta^*)^2\sim t^*=1/16,\nu^*\sim\pm\sqrt{\epsilon}$ where $t=\eta/(16\pi J)$, indicating that for $t>t^*$, the vortex-driven localization corresponding to BKT transition ($N\rightarrow 1$) occurs at zero vortex fugacity acting as the boundary line between two supposedly distinct insulating phases ($\nu<0$ and $\nu>0$).In our 3d WSM, since $\nu^*$ is of order unity that does not go to zero as $\epsilon\rightarrow 0$, the vortex loop-driven localization phase boundary extends all the way to $t\sim\Delta^2\rightarrow\infty$, and one thus needs finite vortex loop density at any disorder strength for them to proliferate.This partly explains why the actual Anderson localization occurs at relatively strong disorder \cite{Sarma}.This comparison is illustrated in Fig.2 \cite{NotePhaseDiagram}.The observation that in 3d vortex loops need a finite fugacity to proliferate at any disorder strength does not mean that the vortex loops never proliferate.They do proliferate and the surprising finding of this work is that, contrary to naive expectation that topological defect excitations would provide additional impetus for localization, their Berry phase in fact impedes the localization.The relevance of such nonperturbative effect in the short-distance physics is further enforced by the fact that small size vortex loops are energetically favorable (a vortex loop's free energy increases with its diameter as $F\sim (a_L/a_c) \mathrm{log}(a_L/a_c)$ where $a_L,a_c$ are the loop diameter and core size respectively) and thus mandate their inclusion in theory at those short length scales as the remaining force against Anderson localization. 

To determine the contribution of such vortex loop Berry phase to the (de)localization physics quantitatively, we compute the renormalization of the longitudinal conductance $g_{xx}$ due to the change of length scale as well as the Berry phase effect.Following the analyses presented in \cite{LevineLibbyPruisken}\cite{Pruiskendiluteinstanton}, we obtain the RG equation for $g_{xx}$ including the vortex loop Berry phase effect

\begin{equation}\label{RGlongitudconductivitynonperturbative}
\frac{dg_{xx}}{dl}=g_{xx}-{g^0_{xx}}^*-D_0g^3_{xx}\cos\phi_1e^{-4\pi g_{xx}}
\end{equation}  
where $D_0$ is a positive constant that represents the internal energy of a vortex (or anti vortex) loop and we have taken into account, valid in the low density regime, only vortex loops (and anti-vortex loops) with vorticity 1, and crucially $\phi_1$ is the Berry flux of of vortex loop of unit vorticity we have obtained in Eq.(\ref{vortexloopBerryphase}).As we have noted earlier, $\phi_1$ generally takes values that are noninteger multiples of $2\pi$ and thus gives rise to destructive interference between scattering paths.Without loss of generality, we can take $\phi_1=\pi$ to represent such destructive interference.Comparing Eqs. (\ref{RGlongitudconductivityperturbative}) and (\ref{RGlongitudconductivitynonperturbative}), we can see that the last term in Eq.(\ref{RGlongitudconductivitynonperturbative}), which represents vortex loop contribution to the renormalization of longitudinal conductance, produces positive correction to the conductance, which means it tends to resist localization.More precisely, the Anderson localization now occurs at ${g_{xx}}^*={g^0_{xx}}^*+\delta {g_{xx}}^*$ where $\delta {g_{xx}}^*=-D_0{({g^0_{xx}}^*)}^3\exp(-4\pi{g^0_{xx}}^*)/(1+D_0{({g^0_{xx}}^*)}^2\exp(-4\pi{g^0_{xx}}^*)(3-4\pi{g^0_{xx}}^*))<0$, which means that, since the conductance is a decreasing function of disorder strength at strong enough disorder, the Anderson localization occurs at even lower conductance and thus even stronger disorder strength than what one would naively expect from perturbative analysis.Again for comparison, in 2d $\mathbb{Z}_2$ topological band insulator, $\phi_1=0 (\pi)$ for trivial (topological) insulator phase \cite{FuKane}.The vortex Berry phase therefore strengthens the localization from the semimetallic into the trivial insulating state in 2d $\mathbb{Z}_2$ band insulator.Moreover, it is also clear from scaling theory of localization \cite{Gangof4} that 3d electronic systems are more robust against Anderson localization than 2d systems due to the larger phase space for scattering paths in the former.This is reflected by the fact that for 2d electronic systems, an $\mathcal{O}(g_{xx})$ term on the right hand side of RG equation like that in Eqs.(\ref{RGlongitudconductivityperturbative}) and (\ref{RGlongitudconductivitynonperturbative}) is absent \cite{Gangof4}, which readily gives stronger tendency toward localized state.   

\section*{Discussion}

We have discussed the role of nonperturbative effect in the form of topological vortex loop excitations and their Berry phase in the localization physics in the transport property of
disordered 3d time-reversal invariant Weyl semimetals.We have shown that the Berry phase of the vortex loops in the bulk of noncentrosymmetric WSM generally leads to destructive interference between the vortex loops themselves as well as between the particle scattering paths, thus impeding the localization.On the surface (boundary) of the system, even more interesting effects may arise, as one can have surface with magnetic charges of a given sign that are unpaired and are dangling on the surface, being uncompensated.One can clearly expect that the net Berry phase on the surface is nonzero and in fact equals $\pi$ for the ideal case where exactly half of an open vortex line Berry phase ($1/2$ of $2\pi$) takes effect.This would give rise to destructive interference between topological defects and between the particle scattering trajectories on the surface, hindering the Anderson localization and providing another contribution to the robustness of the surface states in WSM, in addition to the topological protection of the surface Fermi arc due to the very existence of two Weyl points of opposite chiralities \cite{HosurQi}.This holds to be the case so long as the bulk is not Anderson localized as otherwise the surface Fermi arc and the bulk chiral anomaly do actually perish.Interestingly, while we were finishing this paper, an absence of chiral anomaly was reported in noncentrosymmetric semimetal NbP \cite{Exp} where spin-orbit coupling is not significant, which can be directly explained by our disorder-driven `chiral symmetry recovery' idea, consistent also with the concurrent occurrence of linear magnetoresistance that was as well predicted to occur in strongly disordered system \cite{MR}.As noted before, our results apply as well to noncentrosymmetric Weyl semimetals with spin-orbit coupling \cite{HuangNatcomm}-\cite{YangNatphys}.

\section*{Methods}

\subsection*{Deduction of the Form of Renormalization Group (RG) Equation for Longitudinal Conductance $g_{xx}$}
The $\beta$ function for the coupling constant $T$ of $d=2+\varepsilon$ nonlinear sigma model $S[Q]=1/(2\pi T)\int d^dr\mathrm{Tr}[(\nabla Q)^2]$ in the orthogonal class AI with coset space $G/H=Sp(m_1+m_2)/(Sp(m_1)\times Sp(m_2))$ is given \cite{Hikami}

\begin{equation}\label{RGnonlinearsigmaortho}
\beta(m_1,m_2,T)=-dT/dl=\varepsilon T -(m_1+m_2-2)T^2-(2m_1m_2-m_1-m_2)T^3+\cdots
\end{equation} 
where $l=\mathrm{log}L$ is the RG scale corresponding to length scale $L$.In the language of transport theory, $\sigma_{xx}=1/T$.In the replica limit $m_1,m_2\rightarrow 0$ and given that $g_{xx}=\sigma_{xx}L$ in $d=3$, one can rewrite the above RG equation in terms of $g_{xx}$ from which one will obtain Eq.(\ref{RGlongitudconductivityperturbative}) in the main text.

\subsection*{Derivation of the Contribution of Nonperturbative Effect to the RG Equation for $g_{xx}$}
The partition function of a system with vortex loop and antivortex loop excitations with vorticities $M^+, M^-$ respectively in the dilute instanton density approximation is given by \cite{Pruisken}
\begin{equation}
Z=\frac{1}{n}\sum_{M^+,M^-}\frac{1}{(M^+)!(M^-)!}(nL^2\sigma^{(0)}_{xx}D_0e^{-4\pi\sigma^{(0)}_{xx}})^{M^++M^-}e^{i\phi_1(M^+-M^-)}
\end{equation}
with positive constant $D_0$ representing the vortex loop internal energy.Considering the $M^+,M^-=0,1$ contributions and resumming the terms by doing re-exponentiation, we obtain the vortex loop Berry phase contribution to the partition function
\begin{equation}\label{partitionfunctionBerry}
Z_{\mathrm{Berry}}=\frac{1}{n}\left(\exp\left(nL^2D_0\sigma^{(0)}_{xx}e^{-4\pi\sigma^{(0)}_{xx}}\cos\phi_1\right)-1\right)
\end{equation}
The vortex loop Berry phase contribution to the free energy is obtained by taking the replica limit $n\rightarrow 0$ on Eq.(\ref{partitionfunctionBerry}) to get $F_{\mathrm{Berry}}=\mathrm{log}Z_{\mathrm{Berry}}=L^2D_0\sigma^{(0)}_{xx}e^{-4\pi\sigma^{(0)}_{xx}}\cos\phi_1$.This vortex loop free energy renormalizes the longitudinal conductivity $\sigma_{xx}$ as $\overline{\sigma}_{xx}=\sigma^{(0)}_{xx}-D_0(\sigma^{(0)}_{xx})^3\cos\phi_1e^{-4\pi\sigma^{(0)}_{xx}}$.Rewriting this contribution in terms of $g_{xx}$ makes use of substitution $\sigma_{xx}=g_{xx}/L=g_{xx}\exp(-l)$ which gives rise to delicate expression.Here, we employ an intuitive argument to simplify the mathematics while preserving the physics qualitatively.As noted before, the free energy of vortex loops increases with their size and thus typically, vortex loops are small sized and remain so as the RG flows.The length scale $L$ in the vortex loop contribution can be associated with this characteristic vortex loop size.Since as the RG flows the $L$ and thus $l=\mathrm{log}L$ increases from $l=0$, we can assign the $l$ in the vortex loop contribution to $l=0$, corresponding to normalized characteristic vortex loop size $L=1$.This, combined with the RG equation derived from Eq.(\ref{RGnonlinearsigmaortho}) discussed above, gives the RG equation (\ref{RGlongitudconductivitynonperturbative}) given in the main text.

\section*{Acknowledgements}

The author thanks Dr. J. Mill\`{e}s for the very helpful discussion on the abstract algebra of matrix groups, Prof. K. Gawedzki, Prof. D. Carpentier and Dr. A. Fedorenko for insightful discussions, Prof. Reza Asgari and Dr. B.Z. Rameshti for the critical reading of the manuscript.

\section*{Author contributions statement}

I.M. conceived the idea, conducted the research and calculations, analysed the results, and wrote the manuscript. 

\section*{Additional information}

More technical details of the calculations are provided in the Supplementary Material.Correspondence and requests for more materials should be addressed to I.M.

\section*{Competing financial interests}
The author declares no competing financial interests.


\begin{thebibliography}{1}
\bibitem{Cancellation}To see this, note that with internode scattering, the action should be written in terms of composite diffuson matrix field $Q(\mathbf{r})$ combining $Q_n(\mathbf{r})$ of the two Weyl points forming the diagonal blocks while off-diagonal blocks come from internode scattering.Given that $T$ and the resulting gauge field $A_i=T^{-1}\partial_i T$ on both cones take configurations belonging to the same topological sector characterized by integer $\mathbb{Z}$ \cite{RyuNJP}, the resulting trace in $S_{CS}[A]$ from $Q(\mathbf{r})$ vanishes.
\end{thebibliography}
\end{document}